\documentclass[onecolumn,12pt,pre]{revtex4}
\usepackage{graphicx}
\usepackage{epstopdf}
\usepackage{graphicx}
\usepackage{epsfig}
\usepackage{subfig}
\usepackage{amsmath}
\usepackage{amssymb}
\usepackage{amscd}
\usepackage{color}
\usepackage{bm}
\newcommand{\fig}[1]{Fig.~\ref{f:#1}}

\newcommand{\bulk}{\text{b}}
\newcommand{\dip}[1]{\bm{\mu_{\text{#1}}}}
\newcommand{\HSS}{hard-spheres~}
\newcommand{\HS}{hard-sphere~}
\newcommand{\NA}{non-additive~}
\newcommand{\rhobC}{\rho^\bulk_{\text{T} }}
\newcommand{\dens}[2]{\rho^{\text{#1}}_{\text{#2}}}
\newcommand{\rhopGC}{\bar{\rho} }
\newcommand{\xbC}{x_2^{\bulk}}
\newcommand{\xbG}{x_2^{\bulk}}
\newcommand{\xpG}{x_2^{\text{p} } }
\newcommand{\xpGC}{\bar{x}_2}
\newcommand{\Enorm}{\bm{E}_\perp}
\newcommand{\Epara}{\bm{E}_\parallel}
\begin{document}

\title{ Mixture of neutral and dipolar-hard spheres confined in a slit pore: field-induced population inversion and demixing}
\author{C. Brunet, J. G. Malherbe and S. Amokrane}
\affiliation{Physique des Liquides et Milieux Complexes, Universit\'{e} Paris-Est (Cr\'{e}teil), Facult\'{e} des Sciences et
Technologie, 61 av. du G\'{e}n\'{e}ral de Gaulle,
94010 Cr\'{e}teil Cedex, France}
\date{30 August 2011}

\begin{abstract}
We study by  Monte Carlo simulation a binary mixture of neutral and dipolar hard-spheres with non-additive diameters. With a view to understanding the interplay between population inversion for an open pore and the demixing phase transitions, the mixture is considered in the bulk and confined between  two parallel hard-walls modeling a slit pore.  A uniform field is applied in the pore in order to control its composition as shown previously. The demixing lines in the bulk and in the pore are studied by the Gibbs Ensemble Monte Carlo method. The open pore-bulk mixture equilibrium is studied by a combination of canonical/grand canonical simulations. A moderate electrostatic coupling is considered for remaining close to the conditions in which a jump in the adsorption of the minority species has been observed at zero field.  Demixing lines are given in the bulk and for two different pore widths in parallel and normal fields, together with population inversion paths. Similarly to the effect of geometrical confinement, a normal field is found to favor the mixed state so that the population inversion does not interfere with demixing. A parallel field leads to more complex scenarios. Some indications for future work are briefly discussed.
\end{abstract}
\maketitle

\section{Introduction}
 Fluid mixtures adsorbed in porous materials have been studied in the past decades with motivations ranging from technological problems to answering basic theoretical questions (see for e.g.  \cite{Fundamentals_Adsorp,Gubbins}). Understanding  the factors that regulate the behavior of the adsorbed fluid such as its composition, the adsorption geometry, the different interactions, etc. is thus of central importance. Among the various mixtures that are studied, colloidal ones play a special role since it is possible to tune to some extent (eg. by adding polymeric depletants) the effective interaction between the particles and their interaction with the confining medium \cite{Bechinger,VanBlaad}. Since colloidal particles are also often polar or polarizable, the combination of geometrical confinement and the action of an external field offers then a unique way to control their behavior in confinement. In recent work, we have  shown  \cite{JCPCharles2,PRE_Charles} this by Monte Carlo simulation for the simple model  of a binary mixture of neutral and dipolar hard-spheres confined between  two parallel hard-walls modeling a slit pore. We have shown that the composition of the adsorbed fluid  can be controlled in a flexible way by applying an external field in the pore, at fixed thermodynamic state of the bulk fluid.  While the effect of an external field on confined polar fluids is well  known (see for example references in  \cite{Klapp_Rev} for theoretical work  and \cite{Marr} for experiments) the originality of the method we have proposed to control the composition is the combination of the demixing instability  in a mixture with unfavorable attraction (or greater repulsion)  between unlike particles  and the lowering of the potential energy of the dipolar fluid in the region where the field is applied. This also differs from the numerous studies in which the adsorption is controlled through some action on the bulk fluid such a  change in pressure. The possibility to control the behavior of the confined fluid  at fixed bulk state  would indeed be a great advantage for some applications. Not having to rely on the subtle combination of specific interactions should also result in a robust method, feasible with simple components. This is why we choose to illustrate  the basic mechanisms \cite{JCPCharles2,PRE_Charles} on the simple mixture of neutral and dipolar hard-spheres and perfectly smooth planar walls. The physical situation closest to this model is then the adsorption of colloids, in  contrast with molecular adsorption in which specific interactions should be considered. The dipolar hard-spheres and the neutral ones interact with a hard-sphere potential with non-additive diameter. The excess repulsion associated with a  positive non-additivity  is  known to favor demixing \cite{Gazzillo}.  This is one of the basic mechanisms for observing the population inversion in the pore-bulk equilibrium.
 The main result we have obtained in refs. \cite{JCPCharles2,PRE_Charles} is the possibility to produce, in the pore, a field induced population inversion when the bulk is near the demixing instability (referred to as the f-PINBI effect): we start from the situation of a pore in equilibrium with a bulk mixture having a tendency to demix and widely different concentrations of the two components;  when the strength of the field applied in the pore region is varied, a jump in the adsorption of the minority component occurs.  After a certain threshold value, the pore becomes  filled with this species, while the former majority component simultaneously desorbs. One may accordingly produce a jump in the physical properties related to the fluid composition by the sole action of the external field.  A similar effect  has been described in refs \cite{Duda2,Kim}, but it was induced by a variation of the bulk density.  Here the thermodynamic state of the bulk fluid is unaffected. 
  
 Both for practical and theoretical reasons, it is important to understand the connection between this phenomenon of population inversion and  the phase transitions in the bulk and in the confined fluid. For specific applications, one may indeed wish to know in advance how to tune the controllable parameters in order to favor the population inversion with respect to the demixing in the pore and vice versa. More fundamentally, the phase diagram of binary mixtures  is known to  be quite complex, even in the simplest case with symmetric interactions - see for instance \cite{Forstmann} for a general discussion and \cite{Pini, Fantoni,Kahl} for explicit calculations on model systems showing for example the progressive substitution of  the evaporation-condensation by demixing as temperature increases.  With dipolar species, the presence of additional parameters (dipole moment, field strength) complicates further the general phase diagram of the mixture, already in the bulk - see for instance refs \cite{Chen,Blair,Range,Szalai2,Almarza,LiLi}. The confinement increases the diversity of possible scenarios, including specific effects such as layering, capillary condensation, etc - see for e.g. \cite{Evans,Bucior,Binder}. This domain is accordingly even less known  than the phase transitions of bulk mixtures. The theoretical study by semi-analytic methods of such inhomogeneous fluids with anisotropic interactions is  accordingly also problematic (see \cite{Szalai}  for a recent  study of confined dipoles).
 
The objective of this work is thus to analyze further this question on the same model as in ref. \cite{JCPCharles2,PRE_Charles}.  We will supplement our previous results relative  to the bulk phase diagram of the non-additive hard-sphere mixture and the population inversion in the presence of dipoles by the phase separation in the bulk and in the pore in the presence of dipoles, first at zero field. Field strength-composition coexistence curves in the pore will then be given for selected values of the total density, with no attempt to do a full mapping of the parameters space. Indeed, even in this simple model, we have at least four physical  parameters: non-additivity parameter $\delta$, pore width, field strength (parallel or normal), dipole moment or temperature,  besides the usual composition variables. It is thus clear than only a limited domain of the parameters space can be explored by simulation in a reasonable time. For the state points considered, we will try to clarify the status of the population inversion with respect to the demixing  transitions in the bulk and in the pore. 

Another question is the choice of the control variables since this determines the
simulation strategy for the  bulk-bulk,  pore-pore and pore-bulk
equilibria. The first one was determined by the standard Gibbs Ensemble  Monte
Carlo method of Panagiotopoulos \cite{Panagiotop} (GEMC).  The pore-pore
equilibrium was studied by the appropriate version \cite{Panagiotop_b} of this
method.  Even if the open pore  for which the PINBI effect occurs
can be studied similarly,  we wish to use as the control variables the total
density and mole fractions  
\textit{in the bulk}. These control variables are indeed those that are easily accessible experimentally.  We thus adopted a strategy which combines a canonical/grand canonical simulation: the chemical potentials are measured in the bulk considered in the canonical ensemble. They are next used to study the fluid in the open pore in the grand canonical (GC) ensemble. 

This paper is organized as follows: In section II, we present the model and detail the simulation methods we used to study it in the different situation (bulk, pore and bulk-pore equilibria). In section III, we present some corresponding  coexistence curves, without field. We then discuss the relation between the field-induced population inversion and the coexistence  for selected values of the total density.  In the conclusion, we summarize the main results and indicate some directions for future work.

\section{Models and simulation techniques }
\subsection{Model}
The model  considered is a binary mixture of neutral hard-spheres (label
$1$) and dipolar hard-spheres (label $2$) of equal diameter
$\sigma$, with a non-additive diameter $\sigma_{12}=\sigma(1+\delta)$.  This parameter is kept fixed to $\delta=0.2$ in this study. The dipolar \HSS bare a point  dipole  $\dip{}$ at their center. In addition to the hard-sphere interaction, the dipoles interact then through the dipolar potential: 
\begin{equation}
U^{ij}_{\text{Dip}}=\frac{\dip{i}\dip{j}}{r_{\text{ij}}^3}-
3\frac{\left(\dip{i}.\bm{r_{\text{ij}}}\right)\left(\dip{j}.\bm{r_{\text{ij}}}
\right)}{r_{\text{ij}}^5}.
\label{e:ULLDip}
\end{equation}

The reduced dipole moment is given by
$\mu^{\ast }=\mu/\sqrt{k_BT \sigma^3}$. For this study, we took  $\mu^{\ast }=1$.  With this magnitude, an energy $\mu^{\ast }E^{\ast }=1$ would correspond for colloids with diameter  $1\mu m$, for example, to $\mu =2\ 10^{5}D$ and  $E^{\ast }=1$ to
$E\approx 6.1\ 10^{-3}V/\mu m$ at $T=300K$.  This relatively weak
value has been chosen so as to remain close to the situation considered in our previous work \cite{JCPCharles2,PRE_Charles}:  when a non-additive mixture of neutral hard-spheres with $\delta=0.2$  is confined in a pore of width $H=3\sigma$ a density induced PINBI occurs for a bulk state $(\rhobC \approx 0.555, \xbC=0.02)$. This non-additivity of the \HS interaction is a distinguishing feature of the model considered here with respect to the mixture of neutral and dipolar \HSS considered in previous studies.
    
 For the purpose of understanding the interplay between the properties of the fluid in the bulk and in confinement  - here the connection between population inversion for an open pore and  the demixing phase transitions in the bulk and in the confined fluid -  this model will be studied in three steps for which we describe below the simulation method used: the bulk-bulk, the pore-pore and  bulk-pore equilibria.

\subsection{Simulation of the bulk fluid }
To study the mixing/demixing transition for the bulk fluid, we performed
standard Gibbs
MC simulation in the $(N_1,N_2,V_T,T)$ ensemble \cite{Panagiotop}. To achieve this, one has to
simulate  two cubic boxes ($A$ and $B$) of fixed total volume $ V_T=V_A+V_B$
with periodic boundary conditions in each direction. The
boxes exchange particles and volume (the box sizes $L_A$ and $L_B$ change during the simulation \cite{Panagiotop}). The
dipolar energy of the periodic system is evaluated by Ewald sums with conducting boundary conditions \cite{Allen}. For the inverse
screening parameter $\alpha$ we took the same value as in our previous work \cite{JCPCharles2,PRE_Charles} $\alpha
=7/L_{A,B}$ The sum in reciprocal space included all lattice vectors
$\textbf{k}=\frac{2\pi\textbf{n}}{L} $, $\textbf{n}=(n_x, n_y,n_z),$ with
$n^2 <80.$

The control variables  are the reduced total density $\dens{b}{T}=(N_1+N_2)\sigma^3/V_T$ and  temperature
defined as $T^{\ast }=1/{\mu^{\ast }}^2$, kept fixed here to $T^{\ast }=1$. To study the coexistence at a given $\dens{b}{T}$, we started the simulations with box lengths ranging from $L_A=L_B= 12 \sigma$  for nearly mixed fluids to $L_A=L_B= 15.5 \sigma$. This  keeps a representative number
of the minority species when the system is nearly completely  demixed. In all cases, we took $N_1=N_2$ particles, a choice appropriate to the moderately asymmetric coexistence curves considered here.  
Each MC cycle  consisted in  $5N_T$  particle  exchanges, $N_T$ particle
translation moves, $N_2$ rotation moves  and $1$ volume exchange. The type  of movement
was chosen at random. We  performed typically
$10000$ equilibration cycles  and from  $2^{15}$ to $2^{17}$ for critical
conditions of accumulation  cycles.  Note that because of the weak dipolar
interaction, dipole exchanges remain efficient. For lower temperature or
higher dipole moment, dipole exchanges become difficult (in this case one might use the identity change proposed by Blair and Patey \cite{Blair} but this was not necessary in the conditions we have investigated).

When the mixture separates, one measures the mean  total
density in each box: ${\dens{b}{A}}= <\frac{(N_1^A+N_2^A).\sigma^3}{V_A}>$ and its
  mean composition in species $i$ : ${\xbG}=<\frac{N_i^A}{N_1^A+N_2^A}>$ (and same for box $B$). Note
  that near the  critical demixing point the phases exchange between the two
  boxes. The equilibrium is then determined from the compositions  histograms.

\subsection{Simulation of the confined fluid}

The fluid is now confined between two  parallel
hard walls  modeling a slit pore of width $H$. Due to the hard-core interaction with the walls,  the centers of the
particles of both species  can move within the interval
$[\sigma/2,H-\sigma/2]$.  An electric field either normal $\bm{E}=E\ \bm{u_z}$ or parallel
$\bm{E}=E\ \bm{u_x}$ to the walls is applied in the pore. To avoid the complication due to image forces,  we assume continuity  of the dielectric constant at the wall boundary. Since this excludes the case of conducting walls, the field is assumed to be generated by some external device, say parallel plates well away from the walls. 
 To take into account the confinement one has to modify the Ewald summation. Periodic boundary conditions  are now applied only
in the  $x$ and $y$ directions. We used the simple modification proposed by Yeh and Berkovitz \cite{Yeh}:  a simulation box of dimensions 
$L_x$, $L_y$, $L_z$ is used in which a vacuum region is 
incorporated in the z-direction such as $L_z=H+L_{\text{vacuum}}=\gamma L$ 
where $L$ is the lateral dimension: $L_x=L_y=L.$ 
Our implementation of this modified 3D-Ewald summation is similar to that
detailed in the paper by Klapp and Schoen \cite{Klapp}. We took for the Ewald
summation the set of parameters: $\alpha= 7/ L$ where $L$ is the
lateral box size, $\gamma=10$ that gives a gap width $\gamma L-H$ and $n^2=80,
$ with  $\textbf{n}=(n_x, n_y,n_z/\gamma),$
for the terms in reciprocal space.   A large gap might be avoided by using the electrostatic layer correction (ELC)  \cite{Arnold,Brodka}.  We did not try this here since our objective was to test the physical ideas rather than to propose a fully optimized algorithm.

To ensure the pore-pore equilibrium, the GEMC simulation must be conducted  as shown in \cite{Panagiotop_b}, the  volume
exchange being replaced by surface exchange. Simulations are thus performed with
two slits of fixed width $H$, fixed total surface $S_A+S_B$ , with
$S_{A,B}=L_{A,B}^2$  and fixed total number of
particles. In the corresponding  $(N_1,N_2,V_T,T)$ Gibbs ensemble $V_T=(S_A+S_B)(H-\sigma)$
is the volume accessible to
 the particles centers, that is $\sigma/2 $ away from the walls.  We thus kept the same control
parameters as for the coexistence in the bulk :
$\dens{p}{T}=(N_1+N_2)\sigma^3/V_T$ (the superscript $p$ being used only to distinguish
the pore from the bulk). However, we do not impose the equilibrium with the
bulk fluid. The first  reason is a practical  one since the GEMC simulation
does not require the knowledge of the chemical potentials. The second one is
that a closed and an open pore are two physically different situations. In
particular, equal densities in the pore and in the bulk does
not imply the equality of the chemical potentials. For an open pore, the
\textit{average} density is imposed by the equality of the chemical potentials
through the exchange of particles with the bulk. If that case, the
correspondence between $\rhobC$ in  the bulk  and the average density $\rhopGC$ in the pore can be done by determining the chemical potentials from a  $(N,V,T)$ simulation in the bulk and a  GC one in the pore.  Finally, the reduced temperature is kept fixed to $T^\ast=1.$ Isothermal paths at varying field strength correspond then to a fixed $\mu^{\ast }$.

As in the bulk,  the mean  total
density in box A for the phase separated confined fluid  is $\dens{p}{A}= <\frac{(N_1^A+N_2^A).\sigma^3}{V_A}>$ and its
  mean composition in species $i$ is $\xpG=<\frac{N_1^A}{N_1^A+N_2^A}>$ (and same for box B).  

For $H=3\sigma$, we start the simulations with  boxes of equal  sizes ranging
from $L_A=L_B= 20\sigma $ to $L_A=L_B= 21.5 \sigma$ for nearly completely demixed fluids. For $H=9\sigma$,  we start with $L_A=L_B=12
\sigma$ and   $L_A=L_B=15
\sigma$  for highly asymmetric composition. Besides these points that are specific to the confinement geometry, 
the dynamics of the simulation (number of cycles, etc.) was the same  as in the bulk studies.  It should be noticed that fluctuations of the volume and of the particle numbers can be significant during the Gibbs ensemble simulation,  due to the limitation of the number of dipoles that can be taken for a reasonable restitution time.  It is thus difficult to accurately determine the coexistence points (see the irregularities in the slopes of the tie lines ) but we have checked that longer runs did not significantly change the coexistence lines.

Finally it should be noticed that  such finite size effects become significant when approaching the critical
point of the mixing/demixing transition. As an example, a symmetric mixture of \NA \HSS  with
$\dens{p}{T}= 0.46$ confined in a pore of width $H=9\sigma$ is found to  be slightly demixed  (with $\xbG=0.46$) for a box size
$L=12\sigma$, whereas it remains mixed (with $\xbG=0.5$) for  $L=18\sigma$. It is difficult to perform such studies with  dipolar \HSS 
due to the number of dipoles to be considered when the boxes are large, especially in confinement.  Since  locating precisely the critical point is not our goal here, we did not investigate this point further. 

\subsection{Fluid confined in an open pore} 
 To establish the connection between the population inversion in a mixture confined in an open pore and the
demixing lines in the bulk and in the pore, we recall first the method used in our previous work to obtain the  population inversion points.  
The pore-bulk mixture equilibrium is determined by the equality of the chemical potentials $\mu
_{1}$ and $\mu _{2}$, in the bulk and in the pore. As in ref. \cite{PRE_Charles} we used for this a combined canonical/grand canonical simulation.  Using a MC simulation in the $(N_{1},N_{2},V,T)$ canonical ensemble, we first measure by Widom's insertion method \cite{Allen}  the chemical potential of both species. The bulk fluid state is then characterized by a total density $\rhobC$ and composition $\xbC$. Note that these overall density  and composition need not necessarily correspond to a homogeneous mixture. The chemical potentials $\mu_{1,2}(\rhobC,\xbC)$ so obtained are next imposed to the slab through a GC simulation from which one measures the mean density
$\rhopGC(\rhobC,\xbC)$ and the mean composition in the pore
$\xpGC(\rhobC,\xbC)$.  In the density-induced PINBI, the population
inversion is characterized by a set of $\xpGC(\rhobC)$ curves, one for each
value of $\xbC$ considered. They describe the change of the mean pore
composition from the bulk one, $\xbC$. For the f-PINBI instead, the inversion
with respect to $\xbC$ is characterized by a set of 
$\xpGC(E^\ast)$ curves, all for the same $\rhobC$.

\section{Results, Discussion}

\subsection{Bulk phase diagram}

\begin{figure}[htbp]
\centering
\includegraphics[clip,width=7.0cm,angle=270]{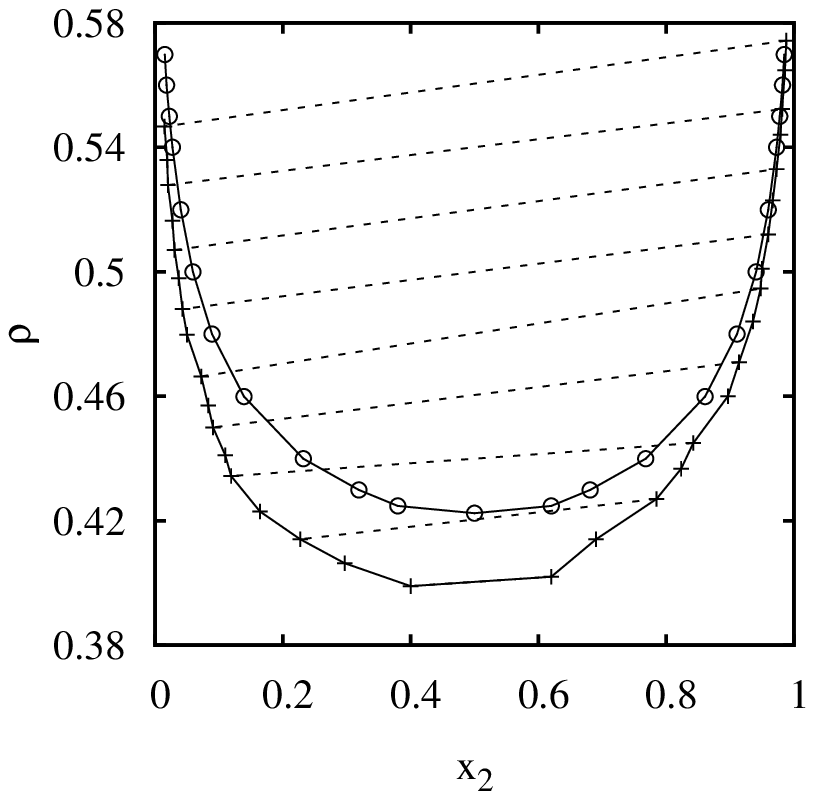}
\caption{Density-concentration demixing line of a bulk mixture of neutral and dipolar non-additive \HSS  with $\delta =0.2$.
Full lines with crosses : $\mu^*=1$, full lines with circles : $\mu^*=0$ (pure \NA \HSS). Tie lines are drawn between some pairs of coexistence points. }
   \label{f:bulk}
\end{figure}

\fig{bulk} compares the bulk phase diagram of a mixture of  neutral and  dipolar
\HSS with that of the \HSS mixture without dipoles. In both cases, the
non-additivity is the same :   $\delta =0.2.$ As expected,
the phase diagram is no longer symmetric with dipolar \HSS: for a fixed total
density  $\dens{b}{T}$, the coexisting phases are a phase rich in dipolar \HSS  and a
poor one, the later with a lower density, as shown by the positive slope of the tie lines. 
As an example, for $\dens{b}{T}=0.51$, 

one has  $\dens{b}{A}=0.498$ in the poor phase ($\xbG=0.037$)
and $\dens{b}{B}=0.523$  in the rich one ($\xbG=0.96$) . The second
observation is that  the lowest  density of demixing (here nearly the critical demixing point) decreases
from $\dens{b}{T}\approx 0.42$ to  $\dens{b}{T}\approx 0.395$ with
the dipoles. Both effects are rather moderate here since we took a low $\mu^*$
for not departing too much from the conditions in which the population
inversion occurs in the pure \HSS mixture \cite{PRE_Charles}. In this respect the
dominant mechanism is here the non-additivity. In comparison, the demixing of
a  mixture of neutral and dipolar hard spheres with \textit{additive} diameters should occur at much
higher density \cite{Blair}, \cite{Chen},  \cite{Szalai2}. For example the demixing  of a mixture with $\xbC=0.2$ occurs \cite{Blair} at $\dens{b}{T}=0.8$ for $\mu^*=1.8$ when we have  $\dens{b}{T}\approx0.418$ already for  $\mu^*=1$. This seems also consistent with the estimation made in ref. \cite {Chen} from integral equations.
Direct comparison with these studies is further complicated by the fact that they used different  $\mu^*$, the pressure instead of the density etc.
Note also that this is not a pure demixing since the density is  not constant \cite{Chen2}.  

With the same parameters of the model, we now consider the mixture confined in
a slab.

\subsection{Fluid confined in a slab of width $H=3 \sigma$}
 \subsubsection{Phase Diagram in confinement \label{H3diag}}

\begin{figure}[htbp]
\centering
\includegraphics[clip,width=7.0cm,angle=270]{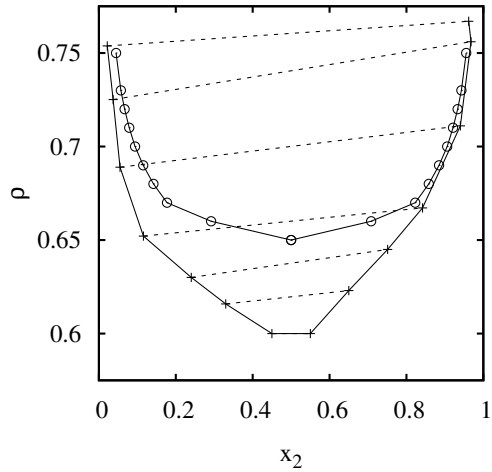}
\caption{Density-concentration demixing line of a mixture confined in a pore of width
  $H=3\sigma$. Meaning of symbols and lines as in \fig{bulk}. }
   \label{f:H3}
\end{figure}

\fig{H3} shows the influence of the confinement on the demixing line. The noteworthy points are: 

- as in the bulk,  the presence of dipoles favors the demixing (the lowest demixing density shifts from
 $\dens{p}{T}\approx 0.65$ to  $\dens{p}{T}\approx 0.595$).

-  as with non-additive hard spheres, the confinement stabilizes the mixture: the lowest demixing 
 density  shifts from $\dens{b}{T}\approx 0.395$ in the bulk to  $\dens{p}{T}\approx 0.595$ for $H=3\sigma$. This effect depends of course on the pore width. Due to the non-additivity of the diameters between the particles and the pure hard core interaction between the latter and the walls, more particles can be packed in the pore than in a bulk slab of same width. This is consistent with the interpretation of the population inversion by an increased free volume in the pore for the minority species \cite {Duda2}. 
For a given width,  the correspondence between a
bulk state point and one in the pore  is established by the equality of the chemical potentials in the bulk-open
pore equilibrium, imposed through the GC simulation. This allows to discuss the position of a state point  with respect to the corresponding demixing lines. For example,  we found that a bulk
state point ($\rhobC=0.51,\xbC=0.02$) corresponds to a single phase in
equilibrium with a single phase also in the pore with
($\rhopGC=0.638,\xpGC=0.06$). Since the bulk mixtures with $\dens{b}{T}=0.51$ demixes at
$\xbG  \approx 0.03$  and the confined one with $\dens{p}{T}=0.638$ at
$\xpG \approx 0.2$, the distance  of the state point ($\rhopGC=0.638,\xpGC=0.06$) to 
the demixing boundary is larger than the corresponding one  in the bulk. This shows that stabilization by confinement
holds also for a mixture confined in an open pore.  

\subsubsection{Connection with the density-induced population inversion}

\begin{figure}[htbp]
\centering
\includegraphics[clip,width=7.0cm,angle=270]{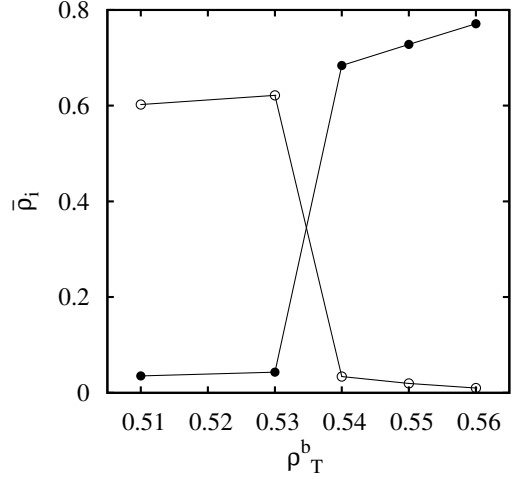}
\caption{Average densities in the confined mixture of neutral and dipolar non-additive \HSS  as a function of the bulk density and fixed bulk concentration $\xbC=0.02$.  Other parameters as in \fig{H3} : $H=3\sigma$, $\delta =0.2$, $\mu^*=1$; \emph{filled circles:}
  adsorption of the dipolar hard-spheres, \emph{empty circles:} desorption of
  the hard-spheres.  The lines are guides to the eye. } 
   \label{f:pinbi}
\end{figure}

\begin{figure}[htbp]
\centering
\subfloat[ Bulk path (demixing line shown in \fig{bulk}) ] {\label{f:trajbulk}\includegraphics[angle=270,origin=br,totalheight=7cm]{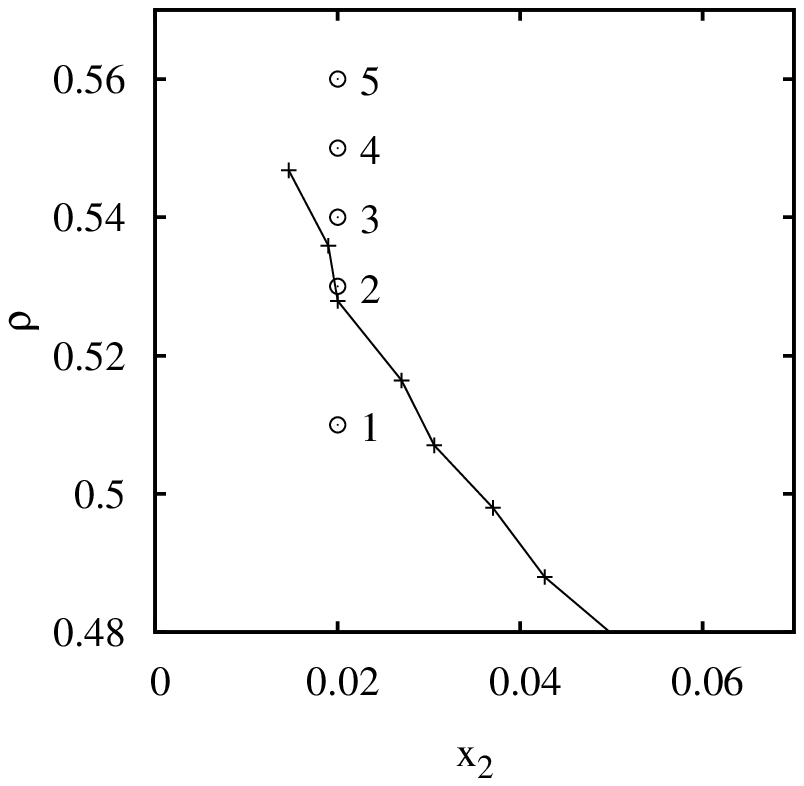}}
\subfloat[Corresponding path in the pore (demixing line shown in \fig{H3}) ]{\label{f:trajslab}\includegraphics[angle=270,origin=br,totalheight=7cm]{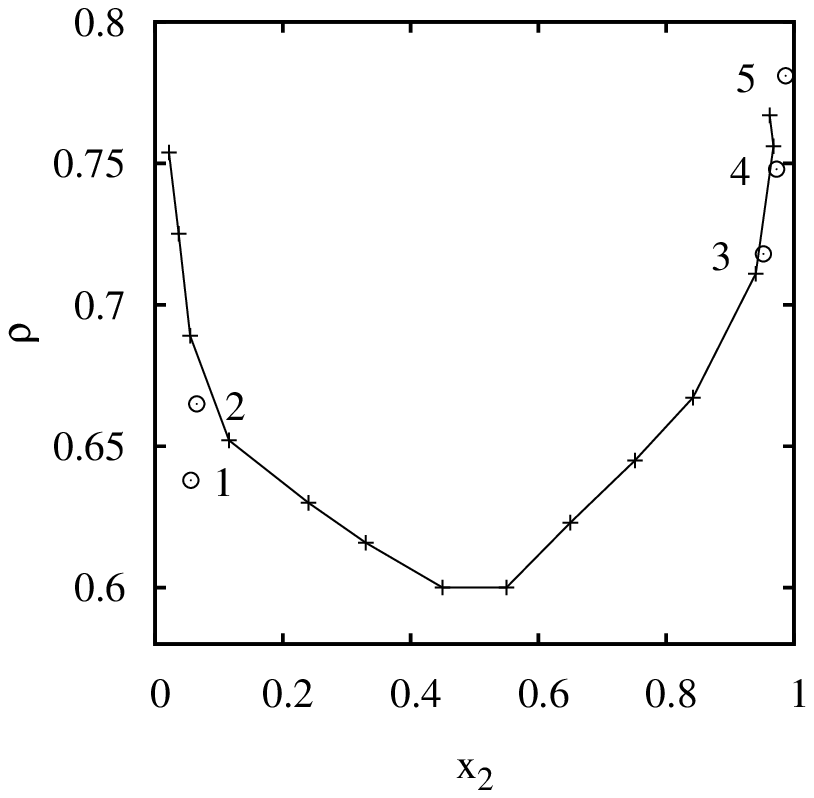}}
\caption{Density-concentration trajectories in the phase diagram  corresponding to the  population
  inversion in  \fig{pinbi}. The bulk composition
  is $\xbC=0.02$. The numbers label corresponding pore/bulk equilibrium state points. }
\label{f:traj}
\end{figure}

 \fig{pinbi} shows the population inversion in a pore in equilibrium with a
 bulk fluid at fixed dipolar concentration  $\xbC=0.02$ and varying
 density. This figure corrects a slight inaccuracy in the similar plots  in
 our previous papers \cite{JCPCharles2,PRE_Charles}: after longer simulations
 the inversion is found to take place between $\rhobC=0.53$ and $\rhobC=0.54$
 and not right after $\rhobC=0.54$.  Comparison with the case of pure hard-spheres (figure 5 in ref. \cite{PRE_Charles}) shows that the influence of the dipolar interaction is moderate. The starting point of the study of the field effect was indeed that the dipolar interaction should not disturb too much the density driven PINBI.   
In ref. \cite{PRE_Charles}, we have shown that with pure non-additive \HSS, the population
inversion occurs when the
bulk fluid starts to become unstable (bulk state points slightly inside
the demixed phase). We confirm here this behavior when the non-additive
mixture comprises dipolar species: as illustrated in \fig{trajbulk}, where we show
the path followed in the bulk fluid at the  fixed composition $\xbC=0.02$, the  population
inversion in the pore (points labeled $2$ and $3$ in \fig{trajslab} ) is indeed obtained for bulk state points  (\fig{trajbulk}) slightly inside  the bulk
coexistence line. The state points in \fig{trajslab} are however outside the pore demixing
line. This confirms that for this actual bulk composition, the population inversion is distinct from the phase
separation in the pore. It corresponds to a sudden change in composition without
coexisting phases: before and after the population inversion, there is only one phase in the
pore but its composition jumps abruptly from nearly the bulk composition
$\xbC$  to the symmetric one $1-\xbC$. This can be viewed as the equivalent in mixtures of the phenomenon of capillary condensation of a confined one-component fluid.  Note however that in order to observe the inversion as shown in  \fig{pinbi}, one needs to prepare the bulk fluid in a metastable state, for these values of the physical parameters.

\subsubsection{Effect of an external field}

\begin{figure}[htbp]
\centering
\subfloat[Normal field ] {\label{f:H3Eperp}\includegraphics[angle=270,origin=br,
totalheight=7cm]{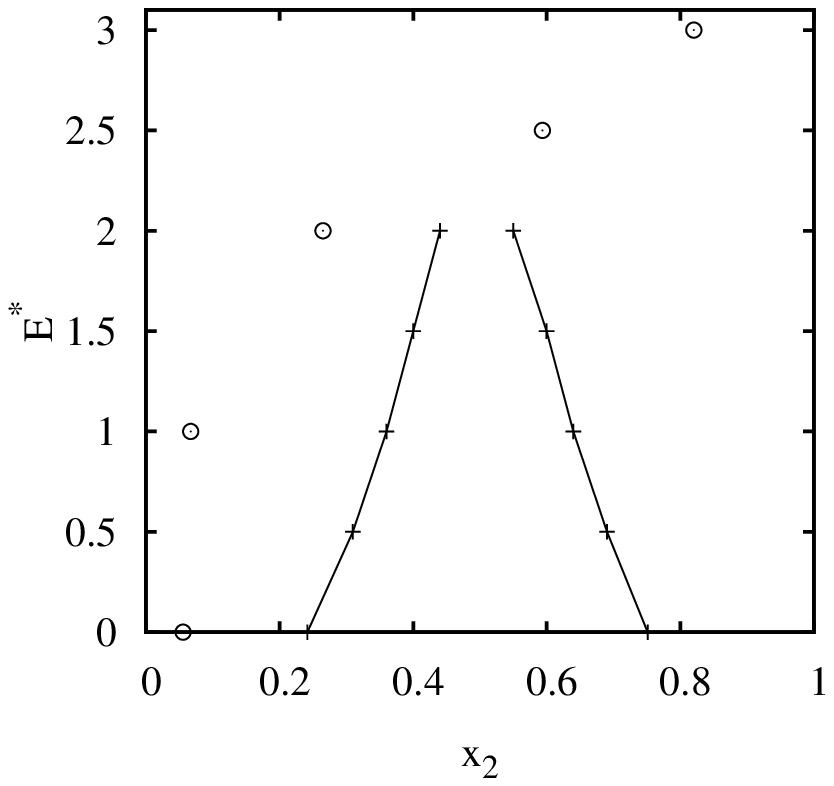}}
\subfloat[Parallel field ]{\label{f:H3Epara}\includegraphics[angle=270,origin=br,totalheight=7cm]{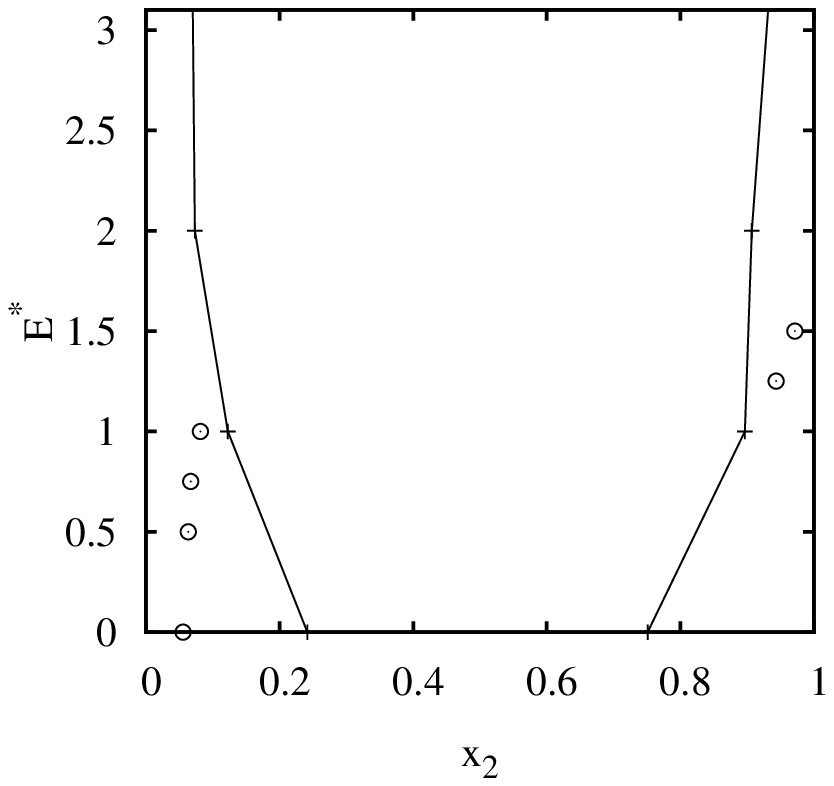}}
\caption{Effect of  the field  strength  on the demixing lines  at fixed total density $\dens{p}{T}=0.638$ in the pore, corresponding for $E^*=0$ to the bulk state $(\dens{b}{T}=0.51,\xbC=0.02)$. The associated f-PINBI path is shown  by symbols. All the parameters are as in \fig{H3}.}  
\label{f:fieldH3}
\end{figure}

Studying the effect of an external field should require mapping the
$(E^\ast,\xpG,\dens{p}{T})$ surface. Since this would require very lengthy
simulations, we did that for selected values of the density. In Sec. \ref{H3diag}, the value $\dens{p}{T}=0.638$ was considered to illustrate the effect of confinement on coexistence. It corresponds to the reference bulk state  at which an important effect  of the applied field  on the composition  of the confined fluid has been evidenced \cite{JCPCharles2,PRE_Charles}. We thus show in \fig{fieldH3}  the effect of a uniform field (for two
directions) on coexistence, the total density being fixed to  $\dens{p}{T}=0.638$.   

In a normal field ($\Enorm$),  the coexistence points  (\fig{H3Eperp}) get closer at increasing field
strength : starting from well separated zero-field values
$(\dens{p}{A}=0.63,\xpG= 0.24)$  and  $ (\dens{p}{B}=0.645,\xpG=0.75)$, the demixed domain shrinks as the field
strength increases. This suggests a critical point beyond which the mixture
becomes homogeneous. 
We also tentatively show f-PINBI points for the
total density $\rhobC=0.51$. They correspond to the bulk composition $\xbC=0.02$ (rightmost curve in figure 9 in \cite{PRE_Charles}):  $(\rhobC=0.51$,$\xbC=0.02)$ corresponds to $(\rhopGC=0.638$, $\xpGC=0.06)$ in the bulk-pore equilibrium at zero-field. The coexistence curve $E^\ast-\xpG$ is for  $\dens{p}{T}$ fixed at this value $0.638$. It corresponds to a cut at  the \textit{constant} density $\dens{p}{T}=0.638$ of  the
$(E^\ast,\xpG,\dens{p}{T})$ surface. Along  the f-PINBI path now, the mean pore density  $\rhopGC$  changes slightly from
$\rhopGC \approx 0.62$ to $\rhopGC \approx 0.638 $ for $E^\ast \approx 0-3$. It makes thus
sense to show the f-PINBI data on the same plot as  the  $E^\ast-\xpG$ coexistence at
fixed $\dens{p}{T}$. We may then infer from their comparison that the field-induced
population inversion occurs in a single-phase state in the pore for this specific bulk state. This scenario should still hold when the homogeneous bulk is at higher $\xbC$  since the tendency to demix is less pronounced in mixtures more symmetric in composition, a trend also favored by the mixing effect of the applied field.  

With  the parallel field ($\Epara$), the coexistence lines show the opposite
trend as shown in \fig{H3Epara}. First, the demixed domain widens as $E^\ast$  increases. Second, the f-PINBI path crosses the demixing line.  It is however
difficult to draw a conclusion only from the position of the f-PINBI points
since the variation of the  mean pore density is now significant ($\rhopGC
\approx0.638-0.70 $ for $E^\ast \approx0-1.5$).  From a study of the composition
histograms in the GC simulation,  we found that the two points closest to the
coexistence line ($E^\ast=1$ and $1.25$), the histograms are centered about a single
value.  At the intermediate value $E^\ast=1.125,$ the histograms become bivariate. This shows that for $\xbC=0.02$ there can exist a narrow range of field
strengths in which one has a demixed phase rather than a population inversion involving a single phase.

In order to understand  the physical mechanisms, it should be noted that the
main effect of the applied field is to align the dipoles and decrease
the potential energy in the pore by a contribution $-\sum\dip{i}\cdot \bm{E}$. Its combination with the hard-core repulsions, the dipole-dipole interactions  and the entropic effects minimizes the appropriate thermodynamic potential at equilibrium. The dipole-dipole interactions involve both attractive contributions between the dipoles that form a chain and repulsions between parallel chains.  These positive contributions  compete  with the decrease in potential energy due to the field.  
The outcome of this combination is difficult to anticipate and depends on the kind of pore - closed or open - and on the field direction. Quite generally,  the hard-core repulsion with the wall limits the length of the dipolar chains in the normal field and their number in the parallel one.
 
In a closed pore (for which the  $E^\ast-\xpG$ coexistence curves at
fixed $\dens{p}{T}$ are more directly relevant), the question is which situation -  homogeneous or demixed  mixture - leads to a minimum free energy. Starting from a demixed mixture (due to the non-additivity) at zero field, an increase of  $\Enorm$ will favor mixing  because only short dipolar chains can form. This also limits the decrease in entropy. The intra-chain attractive contributions that favor demixing are thus bounded and the interchain repulsion makes less efficient the effective repulsion between the neutral and the dipolar \HSS  that drives the demixing.  As a result a normal field favors the mixed state in the closed pore.
In parallel field, formation of long chains parallel to the walls, is favored. An overall decrease of the potential energy can be anticipated due to these long chains. 	This can be directly checked  for dipoles distributed on a regular lattice (see also fig. 10 in ref \cite{Klapp}). The fact that the field reinforces the trend towards demixing induced by  the non-additivity shows that the dominant effect is the decrease in potential energy (this is consistent with the fact that the dipoles are naturally oriented parallel to the walls already at zero field).
 
In the open pore, the grand potential is minimized also by a change in the number of particles (it varies during  the GC
simulation).  Besides the mechanisms that act in the close pore, the filling of
the pore by the dipoles is then energetically favorable.  Accordingly, more and more dipoles enter the pore. With $\Enorm$,  they form  a homogeneous mixture with the hard-spheres,  as for the closed pore. The population inversion is then progressive, with no interference with demixing.
 With $\Epara$ the filling of the pore will interfere with field-reinforced trend to demixing with a complex combination  of both, including a change of the average density (no pure demixing). For example, in \fig{H3Epara} for the point corresponding to  $E^\ast=1.5$, the mean number of dipoles $\bar {N}_2$ in the open
pore is approximately $10\%$  larger than the corresponding one at fixed density  $\dens{p}{T}=0.638$. This larger number of dipoles that benefit from the decrease of potential energy due to the coupling with the field seems consistent with the preference for the population inversion rather than for the phase separation. For other parameters,  the additional flexibility offered by the particle exchange with the bulk can lead to a great diversity of scenarios. This would be especially true with a finite reservoir as in the mesoscopic GC simulations \cite{Neimark,Kierlik}.  

\subsection{Fluid in a slab of width $H=9 \sigma$}
 \subsubsection{Phase Diagram in the pore}
\begin{figure}[htbp]
\centering
\includegraphics[clip,width=7.0cm,angle=270]{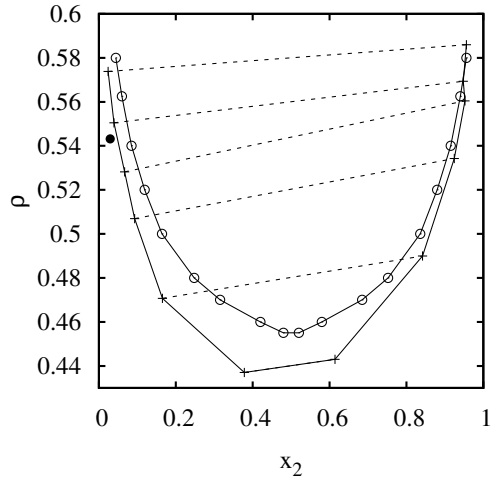}
\caption{Phase diagram of a mixture of neutral and
  dipolar non-additive hard-spheres ($\mu^*=1$) with $\delta=0.2$ confined in a pore of width $H=9\sigma$.\newline
Meaning of the lines and symbols as in \fig{bulk}. The point shown by a circle corresponds to the reference bulk state $(\dens{b}{T}=0.51,\xbC=0.02)$  }

   \label{f:H9}
\end{figure}

\fig{H9} shows the phase diagram for a pore width $H=9\sigma$. The overall effect of
confinement is the same as for $H=3\sigma$, but the situation is quantitatively intermediate between the latter case and the bulk. 
With the dipoles, the lowest density of  demixing  shifts from
$\dens{b}{T} \approx
  0.395$ in the bulk to $\dens{p}{T}\approx 0.435$ in the slab. For further illustrating the lower
  confinement  effect for $H=9\sigma$, we start with the same reference bulk state
  ($\rhobC=0.51,\xbC=0.02$) considered previously.  Keeping  the total  total density in the
  pore fixed to $\dens{p}{T}=0.543$ - that would be the average density (with $\xpGC=0.03$) corresponding to the reference bulk state in the open pore-bulk equilibrium - and varying the dipole concentration $\xpG$,  the mixture demixes into  a dipole poor
  phase with  ($\dens{p}{A}=0.53,\xpG=0.07$) and a rich one with
 ($\dens{p}{B}=0.56,\xpG=0.95)$. At fixed total density $\dens{p}{T}=0.543$,
 the mixture is homogeneous  up to $\xbC\approx 0.05$.  Comparison with the
 discussion in Sec. \ref{H3diag} shows that the stabilizing effect of confinement is less marked than for $H=3\sigma$.

\subsubsection{Effect of the external field}

\begin{figure}[htbp]
\centering
\subfloat[Normal field ] {\label{f:H9Eperp}\includegraphics[angle=270,origin=br,totalheight=7cm]{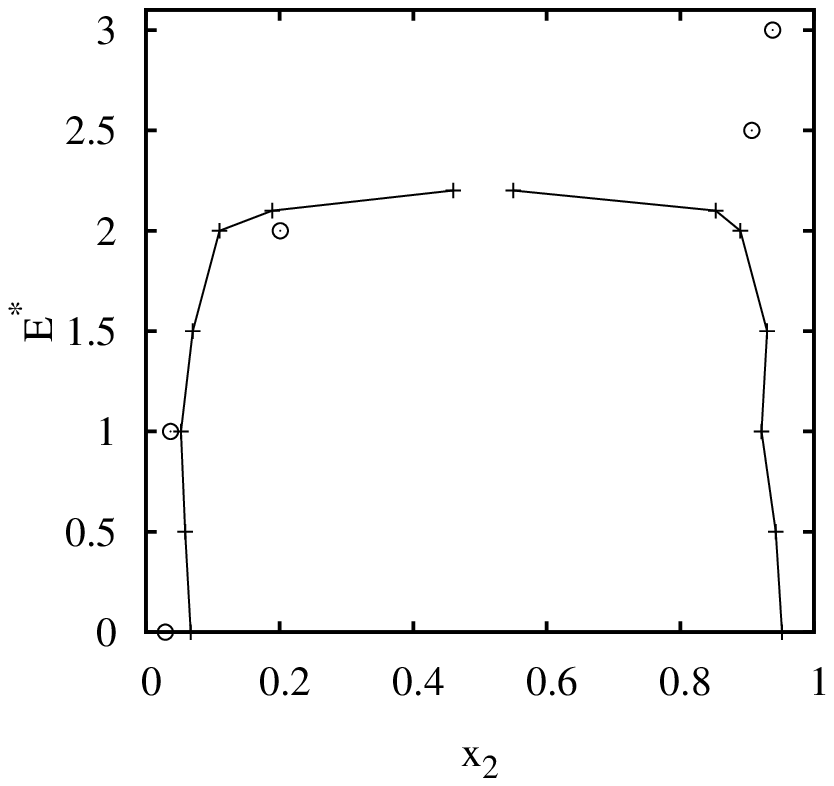}}
\subfloat[Parallel field ]{\label{f:H9Epara}\includegraphics[angle=270,origin=br,totalheight=7cm]{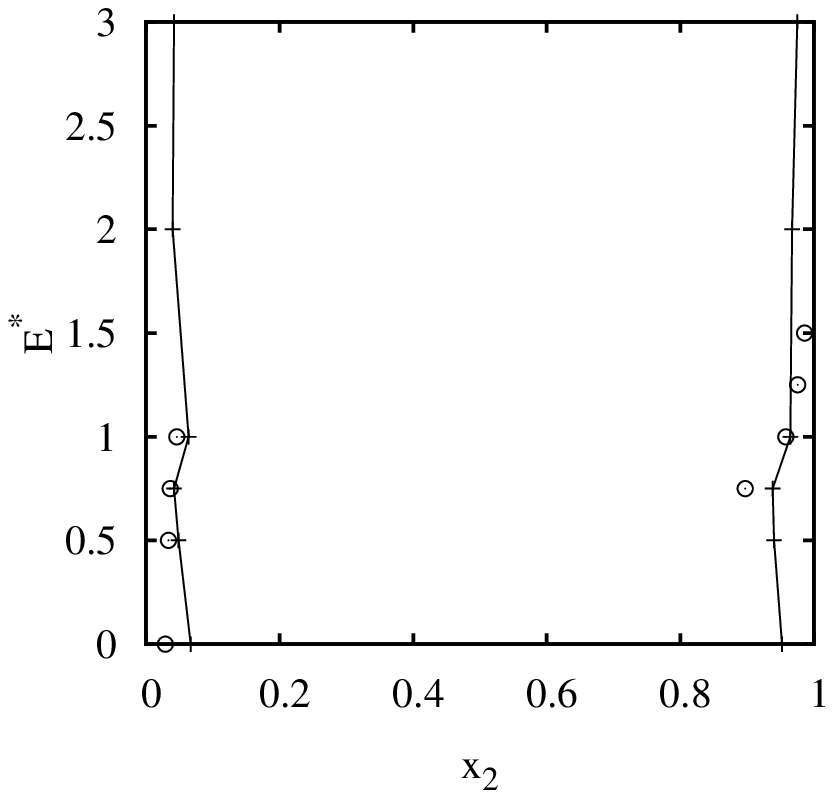}}
\caption{Effect of the field strength on the coexistence line in a pore of width  
$H=9\sigma$ at
  fixed total density $\dens{p}{T}=0.543$. Meaning of the lines and symbols as in  \fig{fieldH3}.}
\label{f:H9.543}
\end{figure}

 Similarly to the case $H=3\sigma$, we start with the value of the total density  $\dens{p}{T}=0.543$ considered above. 
The gross trends found for  $H=3\sigma$ are again observed (\fig{H9.543}) : a high  normal field favors
mixing  while a parallel one favors demixing.
 The f-PINBI points are also shown, with the reservation about the
 representations discussed above. With $\Enorm$, there exists a range of field strengths ($E^*\gtrsim 2$) in which the population inversion goes clearly with a single phase (the PINBI point for $E^*= 2$ which seems inside the demixed phase has an average density actually lower than 0.543). With $\Epara$  the bulk-pore equilibrium leads to coexistence rather than a population inversion with a single phase (see also
 the discussion of the figure 14 in \cite{PRE_Charles}).

\begin{figure}[htbp]
\centering
\subfloat[Normal field ] {\label{f:H9Eperpbis}\includegraphics[angle=270,origin=br,totalheight=7cm]{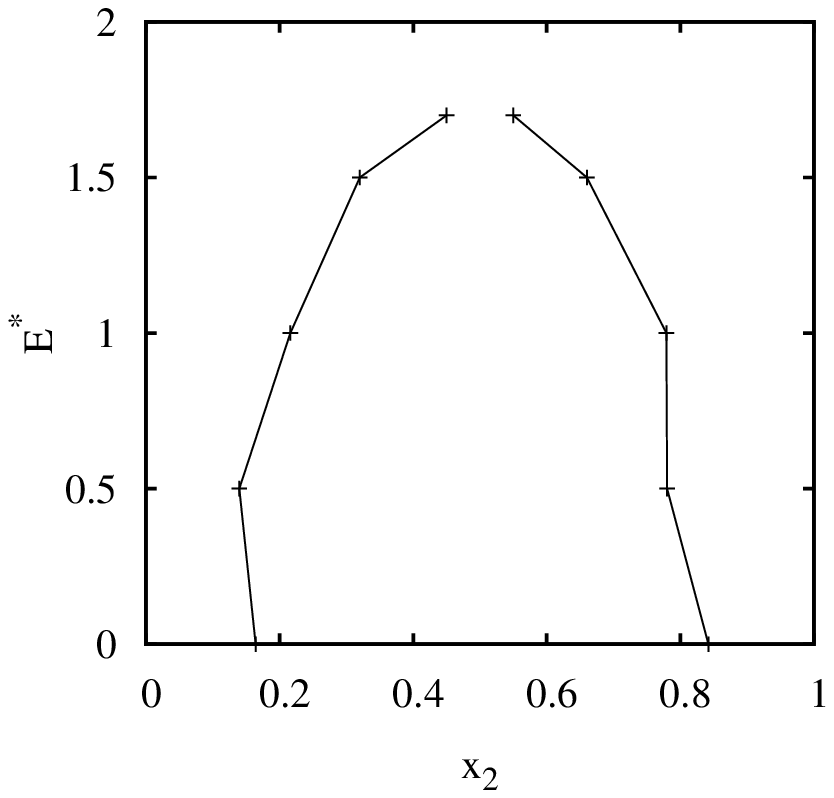}}
\subfloat[Parallel field ]{\label{f:H9Eparabis}\includegraphics[angle=270,origin=br,totalheight=7cm]{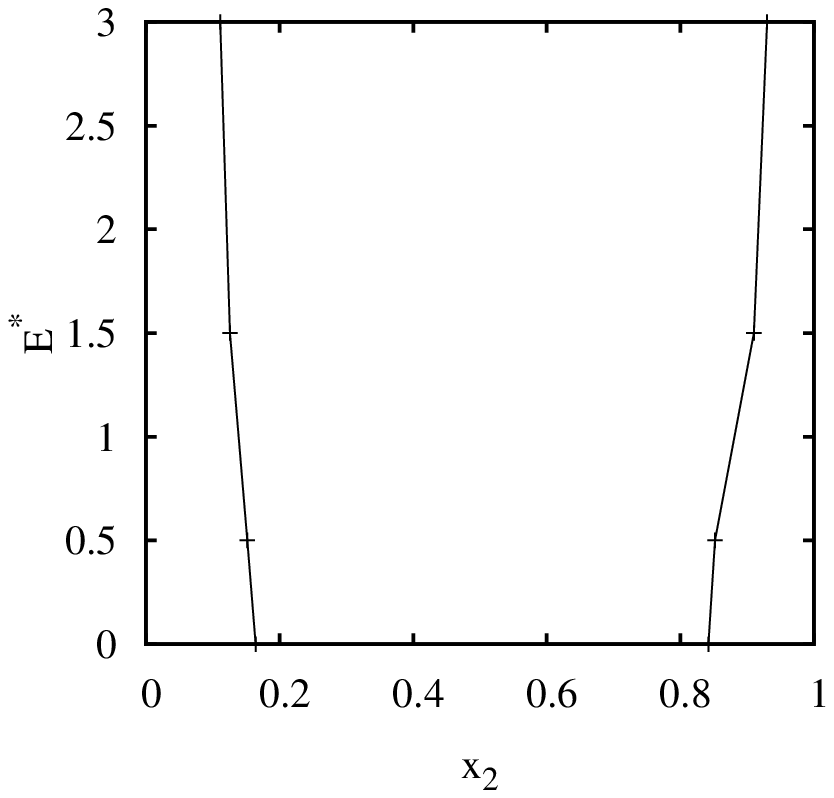}}
\caption{Same as \fig{H9.543} but at 
  fixed total density $\dens{p}{T}=0.48.$}
\label{f:H9.48}
\end{figure}

 As the fluid with a fixed density $\dens{p}{T}=0.543$ is demixed at zero field almost over the entire composition range, the effect of the parallel field on the coexistence  is difficult to see. We therefore considered  a smaller density $\dens{p}{T}=0.48$ (\fig{H9.48}). At $E^\ast=0$, the corresponding
dipolar concentration (in the poor phase),
$\xpG=0.16$,  is comparable to that of the state point for which the field effect was studied  for
$H=3\sigma$. The shape of the coexistence lines are then similar to those for $H=3\sigma$ (\fig{fieldH3}).   
A qualitative difference is  however that  at low field strengths demixing is
favored both for $\Enorm$ and $\Epara$. But a sufficiently high normal field  always
favors mixing in sharp contrast to the effect of a parallel field.

\section{Conclusion}
We may now  summarize the main points of this study. Based on our previous work on a mixture  of non-additive neutral and dipolar hard-spheres in the bulk and confined between hard walls, we have supplemented the results relative to the population inversion in an open pore with the coexistence curves for the bulk and  the confined mixture. A combination of Monte Carlo simulations in the grand canonical ensemble for the pore-bulk equilibrium and in the Gibbs ensemble for the bulk-bulk and pore-pore equilibria made possible the use of the global density and composition of the bulk phase as the control variables. Accordingly,  we have been able to discuss the interplay between population inversion and demixing with the help of these variables that are easily accessible experimentally.  Our main motivation was to analyze further the connection between the phenomenon of population inversion in an open pore subject to an external field  and  the phase transitions in the bulk and in the confined fluid.  Rather than an performing a generic study of the  phase diagram, we considered a single value of the dipole moment, taken  relatively low  precisely for remaining close to the situation in which the phenomenon of  density induced population inversion has been shown in previous studies.  Other choices for the physical parameters or the study of  transitions such as layering, segregation etc.  might require different strategies than the GEMC/GC combination considered here.  

Our main results concern the density driven population inversion  and the field-induced one. At zero field, we observe that  the presence of dipoles favors the demixing as in the bulk,  but the confinement stabilizes the mixture.  The density induced population inversion is then closely related to the bulk instability, as for the  mixture of non-additive hard-spheres without dipoles. 

At non zero field, one may either consider its effect on phase separation at fixed total density, as in a closed pore, and both phase separation and population inversion, in an open one. The behaviors observed depend on the direction of the field.  A normal field favors the mixing, and, for the open pore, a progressive population inversion without	interference with demixing. In parallel field, demixing is favored. The population inversion interferes then with the phase separation. The result is a jump in the concentration of the dipolar species at the threshold field. Both effects are however difficult to separate, especially when the confinement is moderate. 

These observations can be rationalized largely in terms of the combination of  the drop in electrostatic energy in the pore region for the dipolar species,  the attractive intrachain dipolar interaction and the interchain repulsion,  besides the entropy loss due to the alignment of the dipoles. Since these effects depend on the pore width and the direction of the field,  their actual combination is difficult to anticipate.  One should also consider the effect of the neutral hard-spheres whose overall effect is to decrease the electrostatic coupling. It is also difficult to draw general guidelines from the necessarily  limited domain of the space of physical parameters than can be explored by simulation. Our previous studies of more complete versions of this model illustrate some aspects of such simulations \cite{JCPCharles,Charles_MP}. Further studies of this model by semi-analytic methods should thus be useful at least for a gross estimation of the relevant range for these parameters (the same  mixture without dipoles has recently been studied in the group by integral equations  \cite{AbdJPC}) . 

From a practical point of view, we have shown that it is possible to select
bulk states in which the bulk mixture is homogeneous and a field strength for
which the confined mixture is also homogeneous but with an inverted
population. This is precisely one of the important motivations of this study. In particular, the demixing can be avoided by a sufficient confinement as shown by the increase of the lowest demixing density when the pore width decreases. Finally, this analysis of the population inversion and the demixing phase transition has confirmed that a flexible effect of a uniform field can be observed by considering  mixtures  in the regime of high confinement.  This can be paralleled with the case of bulk fluids or dielectric films in which a significant effect can be observed only when the field is inhomogeneous at a macroscopic scale \cite{Tsori_Nature} or the related studies of "electric bottles"\cite{Leunissen}.

\newpage
\section*{Figures and caption}

\begin{thebibliography}{99}

\bibitem{Fundamentals_Adsorp} Fundamentals of adsorption : proceedings of
the Fourth International Conference on Fundamentals of Adsorption, Kyoto,
edited by M. Suzuki,Elsevier Amsterdam (1993).
\bibitem{Gubbins} Z. Tan, and Keith E. Gubbins, J. Phys. Chem. \textbf{96}, 845
(1992).
\bibitem{Bechinger} C. Bechinger,  Current Opinion  in Colloid and Interface Science \textbf{77}, 204 (2002).
\bibitem{VanBlaad}A. Yethiraj, and A. van Blaaderen, Nature (London) \textbf{421}, 513  (2003)
\bibitem{JCPCharles2} C. Brunet, J. G. Malherbe, and S. Amokrane, J. Chem. Phys.  \textbf{131}, 221103 (2009).
\bibitem{PRE_Charles}  C. Brunet, J. G. Malherbe, and S.  Amokrane,
  Phys. Rev. E \textbf{82}, 021504 (2010).
\bibitem{Klapp_Rev} S. H. L. Klapp, J. Phys. Condens. Matter 17, R525 (2005).
\bibitem{Marr} T.  Gong, and D. W. M.  Marr,  Langmuir \textbf{17}, 2301 (2001).
\bibitem{Gazzillo} D. Gazzillo, J. Chem. Phys. \textbf{95}, 4565 (1991).
\bibitem{Duda2} F. Jim\'enez-Angeles, Y. Duda, G. Odriozola, and M.
Lozada-Cassou, J. Phys. Chem. C, \textbf{112}, 18028 (2008).
\bibitem{Kim} S. Kim, S. Suh, and B. Seong, Journal of the Korean Physical
Society \textbf{54}, 60 (2009).
\bibitem{Forstmann} O. Antonevych, F. Forstmann, and E. Diaz-Herrera, Phys. Rev. E \textbf{65}, 061504 (2002).
\bibitem{Pini}D. Pini, M. Tau, A. Parola, and L. Reatto, Phys. Rev. E \textbf{67}, 046116 (2003).
\bibitem{Fantoni} R. Fantoni, D. Gazzillo, and A. Giacometti, Phys. Rev. E  \textbf{72} , 011503   (2005). 
\bibitem{Kahl} J. K\"ofinger, N. B. Wilding, and G. Kahl, J. Chem. Phys. \textbf{125}, 234503 (2006). 

\bibitem{Chen} X. S. Chen, M. Kasch, and F. Forstmann, Phys. Rev. Lett
  \textbf{67}, 2674 (1991);  X.S. Chen, and F. Forstmann, Mol. Phys. \textbf{76}, 1203 (1992)
\bibitem{Blair} M. J. Blair, and G. N. Patey, Phys. Rev. E \textbf{57}, 5682 (1998).
\bibitem{Range} G. Range, and S. Klapp, Phys. Rev E \textbf{70}, 031201 (2004).

\bibitem{Szalai2} I. Szalai, and S. Dietrich, Molec. Phys. \textbf{103}, 2873 (2005).
\bibitem{Almarza} N. G. Almarza, E. Lomba, C. Martin, and A. Gallardo, J. Chem. Phys. \textbf{129}, 234504 (2008). 
\bibitem{LiLi} L. Li, L. Liang-Sheng, and C. Xiao-Song, Commun. Theor. Phy. \textbf{52}, 523 (2009).
\bibitem{Evans} R. Evans, U.M.B. Marconi, and P. Tarazona, J. Chem. Phys. 
\textbf{84}, 2376 (1986) ; A. O. Parry, C. Rasc\'on, N. B. Wilding, and R. Evans, Phys. Rev. Lett. \textbf{98}, 226101 (2007).
\bibitem{Bucior} K. Bucior, A. Patrykiejew, O. Pizio, and S Sokolowski,
  J. Colloid and Interf. Sci. \textbf{259}, 209 (2003). 
\bibitem{Binder}A. Winkler, D. Wilms, P. Virnau, and K. Binder, J. Chem. Phys. \textbf{133},164702 (2010).
\bibitem{Szalai} I. Szalai, and  S. Dietrich, Eur. Phys. J. E \textbf{28}, 347 (2009). 
\bibitem{Panagiotop} A. Z. Panagiotopoulos, Molec. Phys. \textbf{61}, 813
(1987).
\bibitem{Panagiotop_b} A. Z. Panagiotopoulos,  Molec. Phys. \textbf{62}, 701
(1987).
\bibitem{Allen} M. P. Allen and D. J. Tildesley, \emph{Computer simulation
of liquids} (Oxford Science Publication, UK) (1987).
\bibitem{Yeh} I. -C. Yeh, and M. L. Berkowitz, J. Chem. Phys. \textbf{111}, 3155 (1999). 
\bibitem{Klapp} S. H. L. Klapp, and M. Schoen, J. Chem. Phys. \textbf{117}, 8050 (2002).
\bibitem {Arnold} A. Arnold, J. de Joannis, and C. Holm , J. Chem. Phys. \textbf{117}, 2496 (2002) ; J. de Joannis, A. Arnold, and C. Holm, J. Chem. Phys. \textbf{117}, 2503 (2002).
\bibitem{Brodka} A. Brodka, Chem. Phys. Lett. \textbf{400}, 62 (2004).
\bibitem{Chen2} X. S. Chen, and F. Forstmann, J. Chem. Phys. \textbf{97}, 3696
  (1992).
\bibitem{Neimark} A. V. Neimark, and A. Vishnyakov, J. Chem. Phys. \textbf{122}, 234108 (2005).
\bibitem{Kierlik} J. Puibasset, E. Kierlik, and G. Tarjus, J. Chem. Phys. \textbf{131},124123 (2009).
\bibitem{JCPCharles} C. Brunet, J. G. Malherbe, and S. Amokrane, J. Chem.
Phys. \textbf{130}, 134908 (2009).
\bibitem{Charles_MP} C. Brunet, J. G. Malherbe, and S. Amokrane, Molec. Phys. \textbf{108}, 1773 (2010).
\bibitem{AbdJPC} A. Ayadim and S. Amokrane, J. Phys Chem B \textbf{114}, 16824 (2010).
\bibitem{Tsori_Nature} Y. Tsori, F. Tournilhac, and L.  Leibler, Nature \textbf{430}, 544 (2004).
\bibitem{Leunissen}M. E. Leunissen, M. T. Sullivan, P. M. Chaikin, and
  A. van Blaaderen, J. Chem. Phys. \textbf{128} 164508 (2008).
\end{thebibliography}
\end{document}